\documentclass[%
floatfix,
showkeys,
nofootinbib, %
superscriptaddress, %
]{revtex4-1}

\usepackage{cmap}

\usepackage{ucs}
\usepackage[utf8x]{inputenc}
\usepackage[T1,T2A]{fontenc}
\usepackage[german,russian,english]{babel}

\usepackage[sort&compress]{natbib}
\usepackage{amsmath}
\usepackage{amssymb}

\usepackage{mathtools}
\mathtoolsset{
showonlyrefs,
mathic = true
}

\allowdisplaybreaks

\usepackage{hyperref}
\hypersetup{backref,
 colorlinks=false}
\hypersetup{pdfborder=0 0 0}

\usepackage{microtype}
\UseMicrotypeSet[protrusion]{alltext}

\usepackage{graphicx}

\usepackage[scanall]{psfrag}

\usepackage{listings}
\usepackage{listingsutf8}
\usepackage{fixltx2e}

\usepackage{nicefrac}

\makeatletter
\def\ps@pprintTitle{%
     \let\@oddhead\@empty
     \let\@evenhead\@empty
     \let\@oddfoot\@empty
     \let\@evenfoot\@oddfoot}
\makeatother

\usepackage{physics}
\begin{document}

\title{Finslerian representation of the Maxwell equations}

\author{Dmitry S. Kulyabov}
\email{kulyabov-ds@rudn.ru}
\affiliation{Department of Applied Probability and Informatics,\\
  Peoples' Friendship University of Russia (RUDN University),\\
  6 Miklukho-Maklaya St, Moscow, 117198, Russian Federation}
\affiliation{Laboratory of Information Technologies\\
  Joint Institute for Nuclear Research\\
  6 Joliot-Curie, Dubna, Moscow region, 141980, Russia}

\author{Anna V. Korolkova}
\email{korolkova-av@rudn.ru}
\affiliation{Department of Applied Probability and Informatics,\\
  Peoples' Friendship University of Russia (RUDN University),\\
  6 Miklukho-Maklaya St, Moscow, 117198, Russian Federation}

\author{Tatyana R. Velieva}
\email{velieva-tr@rudn.ru}
\affiliation{Department of Applied Probability and Informatics,\\
  Peoples' Friendship University of Russia (RUDN University),\\
  6 Miklukho-Maklaya St, Moscow, 117198, Russian Federation}

\author{Anastasia V. Demidova}
\email{demidova-av@rudn.ru}
\affiliation{Department of Applied Probability and Informatics,\\
  Peoples' Friendship University of Russia (RUDN University),\\
  6 Miklukho-Maklaya St, Moscow, 117198, Russian Federation}

\begin{abstract}
When the Maxwell equations are geometrized, the Maxwell Lagrangian is
usually reduced to the Yang-Mills Lagrangian. In this case, the
effective quadratic metric, usually corresponding to the Riemannian
metric of our space, is considered. However, it is more
reasonable to use Finsler approach to Maxwell's equations. In the paper
the Finsler representation of the geometrized Maxwell equations is considered.
 The comparison with the Riemannian approach also is made.
\end{abstract}


\maketitle

\section{Introduction}
\label{sec:intro}

The ideology of transformational optics is based on the following
trick~\cite{plebanski:1960:electromagnetic_waves,leonhardt:2009:light,pendry:2006:controlling-em,kulyabov:2017:sfm:geometrization_maxwell}.
The Maxwell's equations for medium are usually written for flat Minkowski space.
Also Maxwell's equations may be written in vacuum, but for an arbitrary
Riemannian space. By equating the relevant terms in these
systems, we obtain the geometrization of the medium parameters. However,the given
 method has certain drawbacks. The permeability (permittivity) tensor is
constructed from the metric tensor~\cite{tamm:1925:mathann,tamm:1925:jrpc::en}, 
but the metric tensor in Riemannian space has only 10 independent components. This
not enough to describe the full permittivity tensor.
So it seems justified to use a richer geometric
structure such as Finsler geometry~\cite{kulyabov:2018:sfm:riemannian-not-sufficient}.

The paper presents the basic elements of Finsler
geometry. On this basis, we use differential forms to write
Maxwell's equations both for Riemannian and Finsler
geometries~\cite{lammerzahl:2018:finsler-gravity,itin:2014:finsler_coulomb}.

\section{Notations and conventions}
\label{sec:notation}

  \begin{enumerate}

  \item We will adhere to the following agreements. Greek indices
    ($\alpha$, $\beta$) will refer to the four-dimensional space.  Latin
    indices will refer
    to the space of arbitrary dimension.
    
  \item The comma in the index denotes a partial derivative with respect to
    corresponding coordinate ($f_{, i} := \partial_{i}f$);  the semicolon
    denotes a covariant derivative ($f_{;i} := \nabla_{i} f$).

  \end{enumerate}

\section{Elements of Finsler geometry}
\label{sec:finsler}

Finsler geometry can be considered as geometry without restriction
by quadratic metric~\cite{chern:1996:finsler-quadratic-restriction}. In the general case the Finsler
metric depends not only on the coordinates, but also on the
velocities~\cite{rund:book:finsler::en,asanov:book:finsler::en}.

\subsection{General definitions}

Let $M$ be a smooth $n$-dimensional manifold, $TM$ is a tangent
bundle over $M$, $(x^i)$ are local coordinates
on $M$, $(x^i , y^i)$ are natural local coordinates on $TM$.
The Finsler structure on $M$ is determined by scalar function $L (x, y)$ on
$TM$ which satisfies the following conditions:
\begin{itemize}

\item $L(x, y)$ --- positive homogeneous function of the first degree
  with respect of the tangent vector coordinates:
\begin{equation}
L(x, \lambda y) = \lambda(x, y), \quad \lambda > 0;
\end{equation}

\item $L(x, y)$ is positive if $y \neq 0$:
\begin{equation}
L(x, y) > 0, \quad y \neq 0;
\end{equation}

\item the quadratic form $\varphi(\xi) = g_{ij} (x, y)\xi^i \xi^j$ is
positive definite, that is $\varphi(\xi) > 0$ for all $\xi \neq 0$,
where functions
\begin{equation}
g_{ij} := F_{\cdot i \cdot j} 
\end{equation}
are components of the non-degenerate tensor field $g$  
(the metric tensor of the space). Here 
\begin{equation}
  F_{\cdot i} := \pdv{F}{y^i}.
\end{equation}
\end{itemize}

Manifold $M$ with given Finsler structure $L$
is called the Finsler space $F^n$. The function
\begin{equation}
F = L^2
\end{equation}
is called the metric function of the Finsler space $F^n$.
If $x$ and $x + \dd{x}$ are two infinitely close points of space
$F^n$, then the distance $\dd{s}$ between them  is the value of function $L$
in point $(x, \dd{x})$:
\begin{equation}
\dd{s} = L(x, \dd{x}).
\end{equation}

The length $s$ of the curve $c: x = x (t)$ that connects the points $x_1 = x (t_1)$ and
$x_2 = x (t_2)$ is determined by the integral
\begin{equation}
s = \int_{t_1}^{t_2} L(x(t), \Dot{x}(t)) \dd{t},
\end{equation}
where $\Dot{x}(t) = \dv{x (t)}{t}$ is
the velocity vector of the curve.
The fundamental function $F$ is a homogeneous function of the second
degree from the coordinates of the tangent vector. Therefore, by the Euler theorem
we get: 
\begin{equation}
y^i F_{\cdot i} = 2F.
\end{equation}
Differentiating this expression by $y^I$, we obtain:
\begin{equation}
y^i F_{\cdot i\cdot j} = F_{\cdot j},
\end{equation}
from where, from convolution with $y^j$, we get:
\begin{equation}
g_{ij} y^i y^j = F
\end{equation}
and
\begin{equation}
\dd{s}^2 = g_{ij} (x, \dd{x}) \dd{x^i} \dd{x^j}.
\end{equation}

In the last expression the functions $g_{ij} (x, y)$ are homogeneous functions of $y$: 
\begin{equation}
y^k g_{ij\cdot k} = 0.
\end{equation}

In Finsler geometry the characteristic element is the tensor
\begin{equation}
C_{ijk} =
\frac{1}{2}
g_{ij\cdot k},
\end{equation}
the vanishing of which is a necessary and sufficient condition for
 the space $F^n$ to be Riemannian. This tensor is symmetric
by all indices. 

In the tangent space $T_{x}M$ we will consider the equation
\begin{equation}
L(x^i , y^i ) = 1.
\end{equation}

By identifying $T_{x}M$ with centroaffine space, the center 
$o$ of which $(y^i = 0)$ is the tangent point $x \in M$, we can consider 
that this equation defines in $T_{x} M$ a hypersurface which is called as the indicatrix. 
It turns out that 
points $y$ of $T_{x} M$ satisfying the inequality
\begin{equation}
L (x^i , y^i ) \leqslant 1,
\end{equation}
are internal or boundary points of a convex
body whose boundary is given by the indicatrix equation. This is
the consequence from the conditions of the basic definition.
Now we can define the length of the vector $y$ of the centroaffine
space $T_{x} m$ by using equality
\begin{equation}
|y| = L (x^i , y^i).
\end{equation}

Thus defined metric in the vector space $T_{x} M$
will be the Minkowski metric. Therefore, the Finsler space can be
defined as a smooth manifold $M$ and in its tangent spaces
$T_{x} M$ one can define the Minkowski metric which
depends on $x \in M$ as smooth function.

The manifold $M$ is called Finsler space
$F^n$ with alternating metric if in its tangent bundle $T M$
the metric is given by homogeneous function $F (x, y)$ of second degree
on the coordinates of the tangent vector:
\begin{equation}
F (x, \lambda y) = \lambda^2 F (x, y), \quad \lambda  \neq 0.
\end{equation}
Moreover, it is a non-degenerate function:
\begin{equation}
\det \|F_{\cdot i\cdot j} \| \neq 0.
\end{equation}

\subsection{Berwald сonnection}

In Finsler geometry there are two main connections: the Berwald and the Cartan ones.

The characteristic feature of the Berwald connection is the coincidence of its
geodesic with the extremals of the functional:
\begin{equation}
F (x, \Dot{x})\dd{t}.
\end{equation}

The Euler--Lagrange equations of this functional can be led
to the canonical form due the
nondegeneracy of the metric function $F$:
\begin{equation}
\Ddot{x}^k + 2 G^k (x, \Dot{x}) = 0,
\end{equation}
where
\begin{equation}
G^k := 
\frac{1}{4}
g^{ik} (\partial_j F_{\cdot i} \Dot{x}^j - \partial_i F ).
\end{equation}
Here $g^{ik}$ are contravariant components of the metric tensor
$g_{ik} g^{kj} = \delta_{i}^{j}$. The function $G^k$ is changed according to the following law: 
\begin{equation}
G^{k'} =
\partial_{k} x^{k'} G^{k}
− \frac{1}{2}
\partial^2_{ij}
x^{k'} y^{i} y^{j}.
\end{equation}
Herewith 
$G^{k}_{ij}: = G^{k}_{\cdot i\cdot j}$ are converted as connection coefficients:
\begin{equation}
G^{k'}_{i'j'}  = G^{k}_{ij} \partial_{i'} x^i \partial_{j'}  x^j \partial_k x^{k'} + \partial_p x^{k'} \partial^2_{i' j'}  x^p,
\end{equation}
which determine the Berwald connection. Since 
$G^k (x, y)$ are homogeneous functions of the second degree by $y$, then
\begin{equation}
G^{k}_{ij} y^i y^j = 2 G^k .
\end{equation}

The extremal equations coincide with the geodesic equations in
Berwald connections:
\begin{equation}
\Ddot{x}^k + G^k_{ij} \Dot{x}^i \Dot{x}^j = 0.
\end{equation}

We define the vector field on the tangent bundle $T M$ of the Finsler space $F^n$
\begin{equation}
X = y^i \pdv{}{x^i}
− 2 G^k \pdv{}{y^i}.
\end{equation}
The integral curves $(x (t), \Dot{x} (t))$ determine the geodesic of the Finsler space: 
\begin{equation}
\Dot{x}^k = y^k, \quad \Dot{y}^k = −2 G^k (x, \Dot{x}).
\end{equation}

Berwald connection generates the infinitesimal connection,
that is, the distribution $H: z \to H z$  on
tangent bundle $T M$ of the basis manifold
$M$. At each point $z \in T M$ vectors
\begin{equation}
\delta_i = \partial_i - N_i^k \Dot{\partial}_k
\end{equation}
form the basis of the horizontal distribution, where $N_i^k =
G^k_{\cdot i} := G^k_{ij} y^j$ are the coefficients of the infinitesimal
connectivities,
and $\Dot{\partial}_k: = \pdv{}{y^k}$ is the basis of the vertical
distribution $V : z \to V z$ formed by all vectors 
tangent to the layer at $z$. The dual basis is
$\delta x^i = {\dd{x}^i, \delta y^i }$, where $\delta y^i =
\dd{y}^i + N_{k}^{I} \dd{x}^k$. 

\subsection{Cartan сonnection}

The metric tensor of the Finsler space is not covariantly constant in
Berwald's connectivity. One can specify the connectivity that is
consistent with the metric
(when the metric tensor is covariantly constant).
This is the Cartan сonnection.

Suppose that the infinitesimal connection is given on $T M $.

The connection coefficients $(F_{ij}^k, C_{ij}^k)$ of the
$\nabla$ are defined by expansions
\begin{equation}
\nabla_i \partial_j = \nabla_{\delta_i} \partial j =
F_{ij}^k \partial_k,
\quad 
\Dot{\nabla}_i \partial_j = \nabla_{\Dot{\delta}_i} \partial j =
C_{ij}^k \partial_k.
\end{equation}

We require consistency with the metric:
\begin{equation}
  F_{ij}^k = F_{ji}^k,
  \quad
  C_{ij}^k = C_{ji}^k,
  \quad
  \nabla_{\mathcal{A}} g_{ij} = 0.
\end{equation}

Then 
\begin{gather}
  F_{ij}^k =
  \frac{1}{2}
  g^{ks} \qty(\delta_i g_{sj} + \delta_{j} g_{is} − \delta_{s}
  g_{ij}),
  \\
  C_{ij}^k =
  \frac{1}{2}
  g^{ks} \qty( \Dot{\partial}_i g_{sj} + \Dot{\partial}_j g_{is} -
  \Dot{\partial}_s g_{ij} ).
\end{gather}

The Finsler connection is called the Cartan connection, if $N_i^k =
F_{ij}^k y^j$. The coefficients of the Cartan connection are denoted as
$\Gamma^{*k}_{ij}$:
\begin{equation}
\Gamma^{*k}_{ij} =
\qty{\genfrac{}{}{0pt}{}{k}{ij}}
- \frac{1}{2}
g^{kp}
\qty(g_{pi\cdot s} G^s_{\cdot j} + g_{jp\cdot s} G^s_{\cdot i} − g_{ij\cdot s} G^s_{\cdot p} ).
\end{equation}

The tensor part of the connection $\nabla$ takes the form
\begin{equation}
C^k_{ij}
=
\frac{1}{2}
g^{ks} g_{ij\cdot s }.
\end{equation}

\section{Maxwell equations}
\label{sec:maxwell}

We will write Maxwell equations with use of differential forms~\cite{cartan-henry:book:differential-forms::en,bourbaki:book:algebra1::en}.

\begin{align}
  \dd{F} &= 0,
  \label{eq:maxwell:ext:dF=0}
  \\
  \dd {}^*F &= \frac{4\pi}{c} {}^*j.
  \label{eq:maxwell:ext:d*F=*j}
\end{align}

The 2-form of the electromagnetic field $F$ is expressed in 1-form
field potential $a$
as follows:
\begin{equation}
\label{eq:maxwell:ext:F=dA}
F = \dd{A}.
\end{equation}

In this case, the vector potential $A$ makes sense of connectivity, and the Maxwell tensor $F$
makes sense curvature.

\subsection{Maxwell equation in Riemann geometry}

1-form of the potential of the field is written as follows:
\begin{equation}
A =A_{\alpha }\dd{x^{\alpha }}.
\end{equation}

Assuming that $a = A(x^i)$, we may write the 2-form of the electromagnetic field~\eqref{eq:maxwell:ext:F=dA}:
\begin{equation}
\label{eq:maxwell:riman:F=dA:ext}
F = \frac{1}{2}
F_{\alpha \beta} \dd{x}^{\alpha } \wedge \dd{x}^{\beta }.
\end{equation}
The tensor $F_{\alpha \beta}$ has the form:
\begin{equation}
\label{eq:maxwell:riman:F=dA}
F_{\alpha \beta} = \nabla_{\alpha} A_{\beta} - \nabla_{\beta} A_{\alpha} =
A_{\beta ;\alpha} - A_{\alpha ;\beta} =
A_{\beta ,\alpha} - A_{\alpha ,\beta}.
\end{equation}

The equation~\eqref{eq:maxwell:ext:dF=0} takes the form:
\begin{equation}
  \label{eq:maxwell:riman:dF=0}
  \dd{F} = \dd\dd{A} =
  \frac{1}{2}
  \nabla_{\gamma} F_{\alpha \beta } \dd{x}^{\gamma} \wedge
  \dd{x}^{\alpha } \wedge \dd{x}^{\beta}
  =
  \frac{1}{6}
  \qty(\nabla_{\gamma} F_{\alpha \beta }
  + \nabla_{\alpha} F_{\beta \gamma }
  + \nabla_{\beta} F_{\gamma \alpha })
  \dd{x}^{\alpha } \wedge
  \dd{x}^{\beta} \wedge
  \dd{x}^{\gamma} = 0
\end{equation}
or
\begin{equation}
  \label{eq:maxwell:riman:dF=0:F}
  F_{\alpha \beta ;\gamma}
  + F_{\beta \gamma ;\alpha}
  + F_{\gamma \alpha ;\beta} = 0.
\end{equation}

The second Maxwell equation~\eqref{eq:maxwell:ext:d*F=*j} takes the form:
\begin{equation}
  \label{eq:maxwell:riman:d*F=*j}
  \nabla_{\alpha} F^{\alpha \beta} =
  \frac{4\pi }{c} j^{\beta}
\end{equation}
or in partial derivatives~\cite{kulyabov:2012:vestnik:2012-1}:
\begin{equation}
  \label{eq:maxwell:riman:d*F=*j:partial}
  \frac{1}{\sqrt{-g}}
  \qty[
  \partial_{\alpha}
  \qty(\sqrt{-g} F^{\alpha \beta})
  ]
  =
  \frac{4\pi}{c} j^{\beta}.
\end{equation}

\subsection{Maxwell equation in Finsler geometry}

In the case of Finsler geometry, the natural local coordinates on
$TM$ have the form $(x^{\alpha}, y^{\alpha})$. Then for the metric tensor the following may be written:
\begin{equation}
g_{\alpha \beta} = g_{\alpha \beta} (x^{\delta}, y^{\delta}).
\end{equation}

Accordingly, for the potential vector we have:
\begin{equation}
A_{\alpha} = A_{\alpha} (x^{\delta}, y^{\delta}).
\end{equation}

Let us write down the 2-form of the electromagnetic field~\eqref{eq:maxwell:ext:F=dA}:
\begin{equation}
\label{eq:maxwell:finsler:F=dA:ext}
F = \frac{1}{2}
F_{\alpha \beta} \dd{x}^{\alpha } \wedge \dd{x}^{\beta }
+
F_{\alpha \Bar{\beta}} \dd{x}^{\alpha } \wedge \dd{y}^{\Bar{\beta} }
.
\end{equation}
We expand the last expression in terms of vector potential:
\begin{equation}
  \label{eq:maxwell:finsler:F=dA}
  \begin{gathered}
    F_{\alpha \beta} =
    \delta_{\alpha} A_{\beta} - \delta_{\beta}A_{\alpha}
    =
    A_{\beta ;\alpha} - A_{\alpha ;\beta}
    ,
    \\
    F_{\alpha \Bar{\beta}} = - \partial_{\Bar{\beta}}A_{\alpha}
    =
    - A_{\alpha \cdot \Bar{\beta}}
    .
  \end{gathered}
\end{equation}

We assume that the connection is consistent with the metric. That is,
we will use the Cartan connection.
Then the first Maxwell equation~\eqref{eq:maxwell:ext:dF=0} takes the form:
\begin{equation}
  \label{eq:maxwell:finsler:dF=0:F}
  F_{\Bar{\alpha} \beta ;\gamma}
  + F_{\gamma \Bar{\alpha} ;\beta}
  + F_{\beta \gamma \cdot \Bar{\alpha}}
   = 0.
\end{equation}

The second Maxwell equation~\eqref{eq:maxwell:ext:d*F=*j} takes the form:
\begin{equation}
  \label{eq:maxwell:riman:d*F=*j:partial}
  \frac{1}{\sqrt{-g}}
  \qty[
  \delta_{\beta}
  \qty(\sqrt{-g} F^{\alpha \beta})
  +
  \partial_{\Bar{\beta}}
  \qty(\sqrt{-g} F^{\alpha \Bar{\beta}})
  ]
  =
  - \frac{4\pi}{c} j^{\alpha}.
\end{equation}

We have written down the Maxwell equations in the case of Finsler
geometry. Obviously, the correspondence principle is fulfilled for
the written formulas. That is, when there is a dependence on only
one coordinate $x^i$, we get the case of Riemannian geometry.

\section{Conclusion}
\label{sec:conclusion}

In this paper we make the next step in construction of geometrized
Maxwell's equations based on  Finsler
geometry but not on Riemann geometry. From the viewpoint of authors this will allow not only to construct
a complete permeability tensor, but also to describe an anisotropic medium. As a whole 
this will allow us to develop an adequate method for solving the inverse problem of optics.

\begin{acknowledgments}

The publication has been prepared with the support of the ``RUDN University Program 5-100''
and funded by Russian Foundation for Basic Research (RFBR) according to the research project
No~19-01-00645.

\end{acknowledgments}

 \bibliographystyle{elsarticle-num}

\bibliography{bib/maxwell-geom-finsler/cite}

\end{document}


\title{Финслерово представление уравнений Максвелла}

\author{Д. С. Кулябов}
\email{kulyabov-ds@rudn.ru}
\affiliation{Кафедра прикладной информатики и теории вероятностей,\\
  Российский университет дружбы народов,\\
  117198, Москва, ул. Миклухо-Маклая, д. 6}
\affiliation{Лаборатория информационных технологий,\\
  Объединённый институт ядерных исследований,\\
  ул. Жолио-Кюри 6, Дубна, Московская область, Россия, 141980}

\author{А. В. Королькова}
\email{korolkova-av@rudn.ru}
\affiliation{Кафедра прикладной информатики и теории вероятностей,\\
  Российский университет дружбы народов,\\
  117198, Москва, ул. Миклухо-Маклая, д. 6}

\author{Т. Р. Велиева}
\email{velieva-tr@rudn.ru}
\affiliation{Кафедра прикладной информатики и теории вероятностей,\\
  Российский университет дружбы народов,\\
  117198, Москва, ул. Миклухо-Маклая, д. 6}

\author{А. В. Демидова}
\email{demidova-av@rudn.ru}
\affiliation{Кафедра прикладной информатики и теории вероятностей,\\
  Российский университет дружбы народов,\\
  117198, Москва, ул. Миклухо-Маклая, д. 6}

\begin{abstract}

При геометризации уравнений Максвелла лагранжиан Максвелла обычно
сводится к лагранжиану Янга--Миллса. При этом  рассматривается
эффективная квадратичная метрика, обычно соответствующая римановой
метрике нашего пространства. Однако представляется более оправданным
использовать финслеров подход к уравнениям Максвелла. В работе
рассматривается финслерово представление геометризованных уравнений
Максвелла. Проводится сравнение с римановым подходом.

\end{abstract}

  \keywords{уравнения Максвелла, риманова геометрия, финслерова
    геометрия, трансформационная оптика}

\maketitle

\section{Введение}
\label{sec:intro}

Идеология трансформационной оптики базируется на следующем
трюке~\cite{plebanski:1960:electromagnetic_waves,leonhardt:2009:light,pendry:2006:controlling-em,kulyabov:2017:sfm:geometrization_maxwell}.
Записываются уравнения Максвелла в среде, но в плоском пространстве
Минковского.
Также записываются уравнения Максвелла в вакууме, но в произвольном
римановом пространстве. Приравнивая соответствующие члены в этих
системах, получаем геометризацию параметров среды. Однако, данный
метод имеет определённые недостатки. Тензор проницаемостей
конструируется из метрических тензоров~\cite{tamm:1925:mathann,tamm:1925:jrpc::ru}. Но метрический тензор
риманового пространства имеет только 10 независимых компонент. Этого
не хватает для описания полного тензора проницаемостей.
Представляется оправданным переход к более богатой геометрической
структуре, такой как финслерова геометрия~\cite{kulyabov:2018:sfm:riemannian-not-sufficient}.

В работе приводятся основные элементы финслеровой
геометрии. На основании этого из уравнений Максвелла в формализме
внешних форм записываются уравнения Максвелла как в римановой, так и в
финслеровой геометрии~\cite{lammerzahl:2018:finsler-gravity,itin:2014:finsler_coulomb}.

\section{Обозначения и соглашения}
\label{sec:notation}

  \begin{enumerate}

  \item Будем придерживаться следующих соглашений.  Греческие индексы
    ($\alpha$, $\beta$) будут относиться к четырёхмерному
    пространству. Латинские индексы будут применяться для пространства произвольной
    размерности.

  \item Запятой в индексе обозначается частная производная по
    соответствующей координате ($f_{,i} := \partial_{i} f$);
    точкой с запятой --- ковариантная производная ($f_{;i} := \nabla_{i}
    f$).
 
  \end{enumerate}

\section{Элементы финслеровой геометрии}
\label{sec:finsler}

Финслерову геометрию можно рассматривать как геометрию без ограничения
только квадратичной метрикой~\cite{chern:1996:finsler-quadratic-restriction}. В общем случае финслерова метрика
зависит не только от координат, но и от скоростей~\cite{rund:book:finsler::ru,asanov:book:finsler::en}.

\subsection{Общие определения}

Пусть $M$ есть гладкое $n$-мерное многообразие, $TM$~--- касательное
расслоение над $M$, $(x^i)$~--- локальные координаты
на $M$, $(x^i , y^i)$~--- естественные локальные координаты на $TM$.
Финслерова структура на $M$ определяется заданием на
$TM$ скалярной функции $L(x, y)$, удовлетворяющей следующим условиям:
\begin{itemize}

\item $L(x, y)$ --- положительно однородная функция первой степени по
  координатам касательного вектора:
\begin{equation}
L(x, \lambda y) = \lambda(x, y), \quad \lambda > 0;
\end{equation}

\item $L(x, y)$ является положительной функцией, если $y \neq 0$:
\begin{equation}
L(x, y) > 0, \quad y \neq 0;
\end{equation}

\item квадратичная форма $\varphi(\xi) = g_{ij} (x, y)\xi^i \xi^j$ положительно
определена, то есть $\varphi(\xi) > 0$ при всех $\xi \neq 0$, где функции
\begin{equation}
g_{ij} := F_{\cdot i \cdot j} 
\end{equation}
являются компонентами невырожденного тензорного поля $g$ --- 
метрического тензора пространства. Здесь
\begin{equation}
  F_{\cdot i} := \pdv{F}{y^i}.
\end{equation}
\end{itemize}

Многообразие $M$ с заданной финслеровой структурой $L$
называется финслеровым пространством $F^n$. Функцию
\begin{equation}
F = L^2
\end{equation}
называют метрической функцией финслерова пространства $F^n$.
Если $x$ и $x + \dd{x}$ --- две бесконечно близкие точки пространства
$F^n$, то расстояние $\dd{s}$ между ними есть значение $L$
в точке $(x, \dd{x})$:
\begin{equation}
\dd{s} = L(x, \dd{x}).
\end{equation}

Длина $s$ кривой $c : x = x(t)$, соединяющей точки $x_1 = x(t_1)$ и
$x_2 = x(t_2)$, определяется интегралом
\begin{equation}
s = \int_{t_1}^{t_2} L(x(t), \Dot{x}(t)) \dd{t},
\end{equation}
где $\Dot{x}(t) = \dv{x(t)}{t}$~---
вектор скорости кривой.
Фундаментальная функция $F$ является однородной функцией второй
степени по координатам касательного вектора. Поэтому в силу теоремы
Эйлера имеем:
\begin{equation}
y^i F_{\cdot i} = 2F.
\end{equation}
Дифференцируя это выражение по $y^i$, получим:
\begin{equation}
y^i F_{\cdot i\cdot j} = F_{\cdot j},
\end{equation}
откуда, свернув с $y^j$, получим:
\begin{equation}
g_{ij} y^i y^j = F
\end{equation}
и
\begin{equation}
\dd{s}^2 = g_{ij} (x, \dd{x}) \dd{x^i} \dd{x^j}.
\end{equation}

В последнем выражении функции $g_{ij} (x, y)$ являются однородными функциями нулевой степени по $y$:
\begin{equation}
y^k g_{ij\cdot k} = 0.
\end{equation}

В финслеровой геометрии характерным элементом является тензор
\begin{equation}
C_{ijk} =
\frac{1}{2}
g_{ij\cdot k},
\end{equation}
обращение в нуль которого является  необходимым и достаточным условием того,
что пространство $F^n$ будет римановым. Этот тензор симметричен
по всем индексам.

В касательном пространстве $T_{x}M$ рассмотрим уравнение
\begin{equation}
L(x^i , y^i ) = 1.
\end{equation}

Отождествляя $T_{x}M$ с центроаффинным пространством, центр 
$o$ которого $(y^i = 0)$ есть точка касания $x \in M$, можно считать, 
что это уравнение определяет в $T_{x} M$ гиперповерхность, которая 
называется индикатрисой. Оказывается, что 
точки $y$ из $T_{x} M$, удовлетворяющие неравенству
\begin{equation}
L (x^i , y^i ) \leqslant 1,
\end{equation}
являются внутренними или граничными точками выпуклого
тела, граница которого задаётся уравнением индикатрисы. Это является
следствием условий основного определения.
Теперь можно определить длину (норму) вектора $y$  центроаффинного
пространства $T_{x} M$ с помощью равенства
\begin{equation}
|y| = L (x^i , y^i).
\end{equation}

Определённая таким образом  метрика в векторном пространстве $T_{x} M$
будет являться  метрикой Минковского. Поэтому финслерово пространство можно
определить как гладкое многообразие $M$, в касательных пространствах
$T_{x} M$ которого задана метрика Минковского, гладким образом
зависящая от $x \in M$.

Многообразие $M$ называется финслеровым пространством
$F^n$ знакопеременной метрики, если в его касательном расслоении $T M$
задана метрическая однородная функция $F (x, y)$
второй степени по координатам касательного вектора:
\begin{equation}
F (x, \lambda y) = \lambda^2 F (x, y), \quad \lambda  \neq 0.
\end{equation}
Причём это невырожденная функция:
\begin{equation}
\det \|F_{\cdot i\cdot j} \| \neq 0.
\end{equation}

\subsection{Связность Бервальда}

В финслеровой геометрии выделяются две основные связности~---
связность Бервальда и связность Картана.

Характерной особенностью связности Бервальда является совпадение её
геодезических с экстремалями функционала:
\begin{equation}
F (x, \Dot{x})\dd{t}.
\end{equation}

Уравнения Эйлера--Лагранжа этого функционала в силу
невырожденности метрической функции $F$ можно привести
к каноническому виду:
\begin{equation}
\Ddot{x}^k + 2 G^k (x, \Dot{x}) = 0,
\end{equation}
где
\begin{equation}
G^k := 
\frac{1}{4}
g^{ik} (\partial_j F_{\cdot i} \Dot{x}^j - \partial_i F ).
\end{equation}
Здесь $g^{ik}$ --- контравариантные компоненты метрического тензора
$g_{ik} g^{kj} = \delta_{i}^{j}$. При замене локальных координат функции $G^k$
меняются по следующему закону:
\begin{equation}
G^{k'} =
\partial_{k} x^{k'} G^{k}
− \frac{1}{2}
\partial^2_{ij}
x^{k'} y^{i} y^{j}.
\end{equation}
При этом 
$G^{k}_{ij} := G^{k}_{\cdot i\cdot j}$ преобразуются как коэффициенты связности:
\begin{equation}
G^{k'}_{i'j'}  = G^{k}_{ij} \partial_{i'} x^i \partial_{j'}  x^j \partial_k x^{k'} + \partial_p x^{k'} \partial^2_{i' j'}  x^p,
\end{equation}
которые и определяют связность Бервальда. Так как 
$G^k (x, y)$ являются однородными функциями второй степени по $y$, то
\begin{equation}
G^{k}_{ij} y^i y^j = 2 G^k .
\end{equation}

Уравнения экстремалей совпадают с уравнениями геодезических в
связности Бервальда:
\begin{equation}
\Ddot{x}^k + G^k_{ij} \Dot{x}^i \Dot{x}^j = 0.
\end{equation}

На касательном расслоении $T M$ финслерова пространства $F^n$
определено векторное поле
\begin{equation}
X = y^i \pdv{}{x^i}
− 2 G^k \pdv{}{y^i}.
\end{equation}
Его интегральные кривые $(x (t), \Dot{x} (t))$ определяют геодезические финслерова пространства:
\begin{equation}
\Dot{x}^k = y^k, \quad \Dot{y}^k = −2 G^k (x, \Dot{x}).
\end{equation}

Связность Бервальда порождает инфинитезимальную связность,
т.е. распределение $H : z \to H z$ горизонтальных площадок на
касательном расслоении $T M$ базисного многообразия
$M$. В каждой точке $z \in T M$ векторы
\begin{equation}
\delta_i = \partial_i - N_i^k \Dot{\partial}_k
\end{equation}
образуют базис горизонтального распределения, где $N_i^k =
G^k_{\cdot i} := G^k_{ij} y^j$ --- коэффициенты инфинитезимальной
связности,
а $\Dot{\partial}_k := \pdv{}{y^k}$ --- базис вертикального
распределения $V : z \to V z$, образованного всеми векторами,
касательными к слою в точке $z$. Дуальный ему базис имеет
вид $\delta x^i = {\dd{x}^i , \delta y^i }$, где $\delta y^i =
\dd{y}^i + N_{k}^{i} \dd{x}^k$.

\subsection{Связность Картана}

Метрический тензор финслерова пространства не ковариантно постоянен в
связности Бервальда. Можно задать связность, согласованную с метрикой
(когда метрический тензор ковариантно постоянен).
Это связность Картана. 

Предположим, что на $T M$ задана инфинитезимальная связность.
Пусть
$(F_{ij}^k , C_{ij}^k)$ --- коэффициенты
связности $\nabla$, определяемые разложениями
\begin{equation}
\nabla_i \partial_j = \nabla_{\delta_i} \partial j =
F_{ij}^k \partial_k,
\quad 
\Dot{\nabla}_i \partial_j = \nabla_{\Dot{\delta}_i} \partial j =
C_{ij}^k \partial_k.
\end{equation}

Потребуем согласованность с метрикой:
\begin{equation}
  F_{ij}^k = F_{ji}^k,
  \quad
  C_{ij}^k = C_{ji}^k,
  \quad
  \nabla_{\mathcal{A}} g_{ij} = 0.
\end{equation}

Тогда
\begin{gather}
  F_{ij}^k =
  \frac{1}{2}
  g^{ks} \qty(\delta_i g_{sj} + \delta_{j} g_{is} − \delta_{s}
  g_{ij}),
  \\
  C_{ij}^k =
  \frac{1}{2}
  g^{ks} \qty( \Dot{\partial}_i g_{sj} + \Dot{\partial}_j g_{is} -
  \Dot{\partial}_s g_{ij} ).
\end{gather}

Финслерова связность называется связностью Картана, если $N_i^k =
F_{ij}^k y^j$. Коэффициенты связности Картана обозначаются через
$\Gamma^{*k}_{ij}$:
\begin{equation}
\Gamma^{*k}_{ij} =
\qty{\genfrac{}{}{0pt}{}{k}{ij}}
- \frac{1}{2}
g^{kp}
\qty(g_{pi\cdot s} G^s_{\cdot j} + g_{jp\cdot s} G^s_{\cdot i} − g_{ij\cdot s} G^s_{\cdot p} ).
\end{equation}

Тензорная часть связности $\nabla$ примет вид
\begin{equation}
C^k_{ij}
=
\frac{1}{2}
g^{ks} g_{ij\cdot s }.
\end{equation}

\section{Уравнения Максвелла}
\label{sec:maxwell}

Запишем уравнения Максвелла с помощью внешних (дифференциальных) форм~\cite{cartan-henry:book:differential-forms::ru,bourbaki:book:algebra1::en}.

\begin{align}
  \dd{F} &= 0,
  \label{eq:maxwell:ext:dF=0}
  \\
  \dd {}^*F &= \frac{4\pi}{c} {}^*j.
  \label{eq:maxwell:ext:d*F=*j}
\end{align}

2-форма электромагнитного поля $F$ выражается через 1-форму
потенциала поля $A$
следующим образом:
\begin{equation}
\label{eq:maxwell:ext:F=dA}
F = \dd{A}.
\end{equation}

При этом векторный потенциал $A$ имеет смысл связности, а тензор Максвелла $F$
имеет смысл кривизны.

\subsection{Уравнение Максвелла в римановой геометрии}

1-форма потенциала поля записывается следующим образом:
\begin{equation}
A =A_{\alpha }\dd{x^{\alpha }}.
\end{equation}

Считая, что $A = A(x^i)$, запишем 2-форму электромагнитного поля~\eqref{eq:maxwell:ext:F=dA}:
\begin{equation}
\label{eq:maxwell:riman:F=dA:ext}
F = \frac{1}{2}
F_{\alpha \beta} \dd{x}^{\alpha } \wedge \dd{x}^{\beta }.
\end{equation}

Тензор $F_{\alpha \beta}$ при этом имеет вид:
\begin{equation}
\label{eq:maxwell:riman:F=dA}
F_{\alpha \beta} = \nabla_{\alpha} A_{\beta} - \nabla_{\beta} A_{\alpha} =
A_{\beta ;\alpha} - A_{\alpha ;\beta} =
A_{\beta ,\alpha} - A_{\alpha ,\beta}.
\end{equation}

Уравнение~\eqref{eq:maxwell:ext:dF=0} принимает вид:
\begin{equation}
  \label{eq:maxwell:riman:dF=0}
  \dd{F} = \dd\dd{A} =
  \frac{1}{2}
  \nabla_{\gamma} F_{\alpha \beta } \dd{x}^{\gamma} \wedge
  \dd{x}^{\alpha } \wedge \dd{x}^{\beta}
  =
  \frac{1}{6}
  \qty(\nabla_{\gamma} F_{\alpha \beta }
  + \nabla_{\alpha} F_{\beta \gamma }
  + \nabla_{\beta} F_{\gamma \alpha })
  \dd{x}^{\alpha } \wedge
  \dd{x}^{\beta} \wedge
  \dd{x}^{\gamma} = 0
\end{equation}
или
\begin{equation}
  \label{eq:maxwell:riman:dF=0:F}
  F_{\alpha \beta ;\gamma}
  + F_{\beta \gamma ;\alpha}
  + F_{\gamma \alpha ;\beta} = 0.
\end{equation}

Второе уравнение Максвелла~\eqref{eq:maxwell:ext:d*F=*j} принимает вид:
\begin{equation}
  \label{eq:maxwell:riman:d*F=*j}
  \nabla_{\alpha} F^{\alpha \beta} =
  \frac{4\pi }{c} j^{\beta}
\end{equation}
или в частных производных~\cite{kulyabov:2012:vestnik:2012-1}:
\begin{equation}
  \label{eq:maxwell:riman:d*F=*j:partial}
  \frac{1}{\sqrt{-g}}
  \qty[
  \partial_{\alpha}
  \qty(\sqrt{-g} F^{\alpha \beta})
  ]
  =
  \frac{4\pi}{c} j^{\beta}.
\end{equation}

\subsection{Уравнение Максвелла в финслеровой геометрии}

В случае финслеровой геометрии естественные локальные координаты на
$TM$ имеют вид $(x^{\alpha}, y^{\alpha})$. Тогда для метрического тензора можно записать:
\begin{equation}
g_{\alpha \beta} = g_{\alpha \beta} (x^{\delta}, y^{\delta}).
\end{equation}

Соответственно для вектора потенциала имеем:
\begin{equation}
A_{\alpha} = A_{\alpha} (x^{\delta}, y^{\delta}).
\end{equation}

Запишем 2-форму электромагнитного поля~\eqref{eq:maxwell:ext:F=dA}:
\begin{equation}
\label{eq:maxwell:finsler:F=dA:ext}
F = \frac{1}{2}
F_{\alpha \beta} \dd{x}^{\alpha } \wedge \dd{x}^{\beta }
+
F_{\alpha \Bar{\beta}} \dd{x}^{\alpha } \wedge \dd{y}^{\Bar{\beta} }
.
\end{equation}

Распишем последнее выражение через векторный потенциал:
\begin{equation}
  \label{eq:maxwell:finsler:F=dA}
  \begin{gathered}
    F_{\alpha \beta} =
    \delta_{\alpha} A_{\beta} - \delta_{\beta}A_{\alpha}
    =
    A_{\beta ;\alpha} - A_{\alpha ;\beta}
    ,
    \\
    F_{\alpha \Bar{\beta}} = - \partial_{\Bar{\beta}}A_{\alpha}
    =
    - A_{\alpha \cdot \Bar{\beta}}
    .
  \end{gathered}
\end{equation}

Будем считать, что связность согласована с метрикой. То есть, мы будем
использовать связность Картана.  
Тогда первое уравнение Максвелла~\eqref{eq:maxwell:ext:dF=0} принимает вид:
\begin{equation}
  \label{eq:maxwell:finsler:dF=0:F}
  F_{\Bar{\alpha} \beta ;\gamma}
  + F_{\gamma \Bar{\alpha} ;\beta}
  + F_{\beta \gamma \cdot \Bar{\alpha}}
   = 0.
\end{equation}

Второе уравнение Максвелла~\eqref{eq:maxwell:ext:d*F=*j} принимает вид:
\begin{equation}
  \label{eq:maxwell:riman:d*F=*j:partial}
  \frac{1}{\sqrt{-g}}
  \qty[
  \delta_{\beta}
  \qty(\sqrt{-g} F^{\alpha \beta})
  +
  \partial_{\Bar{\beta}}
  \qty(\sqrt{-g} F^{\alpha \Bar{\beta}})
  ]
  =
  - \frac{4\pi}{c} j^{\alpha}.
\end{equation}

Мы записали уравнения Максвелла в случае финслеровой
геометрии. Очевидно, что для записанных формул выполняется принцип
соответствия. То есть, когда есть зависимость только от одной
координаты $x^i$, мы получаем случай римановой геометрии.

\section{Заключение}
\label{sec:conclusion}

В работе сделан очередной шаг по конструированию геометризованных
уравнений Максвелла на основе не римановой, а финслеровой
геометрии. По мнению авторов это позволит не только сконструировать
полный тензор проницаемостей, но и описать анизотропную среду. В совокупности 
это позволит построить адекватный метод решения обратной задачи оптики.

\def\acknowledgmentsname{Благодарности}

\begin{acknowledgments}

Публикация подготовлена при поддержке Программы РУДН «5-100»
и при финансовой поддержке РФФИ в рамках научного проекта
\textnumero~19-01-00645.

\end{acknowledgments}

  \ifthenelse{\boolean{xetex}}
  {\bibliographystyle{ugost2008l}}
  {\bibliographystyle{gost2008l}}

\bibliography{bib/maxwell-geom-finsler/cite}